%% file: main.tex
\documentclass[fleqn,10pt]{wlscirep}
\usepackage[utf8]{inputenc}
\usepackage[T1]{fontenc}
\usepackage[linesnumbered,boxed,ruled]{algorithm2e}
\usepackage{listings}
\usepackage{subfig}
\usepackage{comment}
\usepackage{yfonts,color}
\usepackage[T1]{fontenc}

\usepackage{amsmath}
\usepackage{varioref}

\usepackage[capitalise,noabbrev]{cleveref}

\usepackage{amsfonts}
\usepackage{booktabs}
\usepackage{siunitx}
\usepackage{wrapfig}
\title{A mathematical model of {\em Bacteroides thetaiotaomicron, Methanobrevibacter smithii,} and {\em Eubacterium rectale} interactions in the human gut}

\author[1+*]{Melissa A. Adrian}
\author[1]{Bruce P. Ayati}
\author[2]{Ashutosh K. Mangalam}
\affil[1]{University of Iowa, Department of Mathematics, Iowa City, IA, 52242, USA}
\affil[2]{University of Iowa, Department of Pathology, Iowa City, IA, 52242, USA}

\affil[*]{maadrian@uchicago.edu}
\affil[+]{Current Affiliation: University of Chicago, Department of Statistics, Chicago, IL 60637}

\begin{abstract}
The human gut microbiota is a complex ecosystem that affects a range of human physiology.  In order to explore the dynamics of the human gut microbiota, we used a system of ordinary differential equations to model mathematically the biomass of three microorganism populations: \textit{Bacteroides thetaiotaomicron}, \textit{Eubacterium rectale}, and \textit{Methanobrevibacter smithii}. Additionally, we modeled the concentrations of relevant nutrients necessary to sustain these populations over time. Our model highlights the interactions and the competition among these three species. These three microorganisms were specifically chosen due to the system's end product, butyrate, which is a short chain fatty acid that aids in developing and maintaining the intestinal barrier in the human gut. The basis of our mathematical model assumes the gut is structured such that bacteria and nutrients exit the gut at a rate proportional to its volume, the rate of volumetric flow, and the biomass or concentration of the particular population or nutrient. We performed global sensitivity analyses using Sobol' sensitivities to estimate the relative importance of model parameters on simulation results.
\end{abstract}
\begin{document}

\flushbottom
\maketitle
%
%
\thispagestyle{empty}

\section*{Introduction}

The human gut microbiota is the collection of microorganisms located in the stomach, large intestines, and small intestines, and this system plays an important role in sustaining overall human health. Each individual’s gut composition is unique, and, besides their genetic makeup, their long-term dietary patterns, specifically concerning the types and amounts of carbohydrates, proteins, and fats consumed, affect its composition \cite{David2014,Ji2015}. Specifically, gut microbiota have been shown to be in involved in numerous physiological processes, such as digestion of undigested food (complex starch), development and regulation of immune system, blocking growth of pathogens, and generation of neurotransmitter and vitamins. Thus, any changes in environmental conditions in this gut ecosystem (microbiota) can result in a shift in its composition, which can predispose and/or worsen chronic inflammatory diseases such as inflammatory bowel diseases, multiple sclerosis, rheumatoid arthritis and neurological diseases like Alzheimer's disease, Parkinson disease and autism \cite{PMID:34531642, PMID:29204955, PMID:31753762}. Though the changes in the microbial composition of the human gut microbiota have been shown to be associated with rapid changes in metabolism and overall health, the underlying interactions among species within this microbiota are not yet well understood \cite{Ji2015}. Among all the factors linked with regulation of gut microbiota, diet has emerged as the strongest as it can override genetic influences on microbiota \cite{PMID:28388917, PMID:32231636}. A better understanding of the microbial dynamics, particularly mathematical approaches, may elucidate the role that the gut microbiota plays in human diseases \cite{David2014}. 

Common approaches to analyzing the human gut microbiota for its diversity and composition include, but are not limited to, next-generation sequencing, metatranscriptomics, culturomics, and mass spectrometry analyses \cite{Kumar2019, Ji2015}. These approaches give little insight as to how species interact amongst themselves and with the human host \cite{Ji2015}. Mathematical modeling, however, attempts to answer such questions and can supplement the knowledge gained from these common approaches. Mathematical models can provide evidence to strengthen a hypothesis about these interactions and offer guiding principles for further study of a phenomenon \cite{Oreskes1994}. 

We use ordinary differential equation (ODE)-based modeling as the main tool for our analysis, which tracks information about biomass and concentration levels over time rather than genomic information as in the case of genetically engineered mouse (GEM) models. In this so-called dynamical system approach, we can identify the system's driving parameters and analyze its stability. Despite the advantage of having easily interpretable terms in the model's equations, this approach still has some key drawbacks, specifically the determination of unknown parameters.

The parameters used in biological models like ours include substrate conversion rates, utilization rates, and cellular death rates, which typically can be determined and validated by laboratory experiments. However, some gut microorganisms are unable to be cultivated in a laboratory setting, leaving specific information about these species’ metabolic activity unknown. Additionally, ODE models are often used to track only a small subset of species within a community.  While this is often due to the difficulty of determining a larger number of parameters when the system is expanded \cite{Kumar2019}, focusing on a few subspecies at a time may aid in identifying the dominant relationships in that subsystem.

Despite the impracticality of ODE-based modeling for some very large systems, ODE models paired with other types of modeling, such as agent-based modeling, can provide an insightful understanding of the gut microbiota’s dynamics and interactions \cite{Kumar2019}. As a first step in our modeling efforts with its limitations in mind, we consider an ODE-based modeling approach for a small-scale system of three abundant microorganisms in the human gut microbiota. Our ODE model is based on those of chemostats in that we include inflow and outflow terms in the equations that are missing in batch models, by their nature.  Chemostats are idealized laboratory systems for microbial ecology that have a long history of mathematical modeling behind them \cite{ SmithHalL.andWaltman2008, Harmand2017}.

Similar approaches to ours, such as those for continuously stirred tank reactor (CSTR) models for anaerobic digestion, have been under development for some time. One example of such a model is the IWA Anaerobic Digestion Model No.1 (ADM1), which provides a modeling foundation for generalized biochemical and physio-chemical systems for anaerobic digestion \cite{Batstone2002}. Additionally, Godon et al. 2013\cite{Godon2013} outlines an overview of anaerobic digestion modeling schemes for organ types for a wide variety of animals, humans included. In particular, it is suggested that the stomach can be reasonably represented as a CSTR, and the large intestine can be represented as a series of CSTRs\cite{Godon2013}. More recently, Jegatheesan and Eberl 2020\cite{Jegatheesan2020} created a CSTR model of microbial functional groups and their metabolic products. This model includes a main compartment, referred to as a lumen, in which substrates and microorganisms interact, and a diffusion compartment, referred to a mucus, in which nutrients are uptaken by the host \cite{Jegatheesan2020}. 

We continue this emphasis currently in the literature on the vital role of the inflows and outflows in the human gut ecosystem in our modeling approach. As a first modeling step with our chosen subsystem of the human gut microbiota, we consider a single chemostat, which is a bioreactor subvariant of generalized CSTRs.

Our representation of the human gut microbiota considers a subset of this system, namely the three microorganisms \textit{Bacteroides thetaiotaomicron}, \textit{Eubacterium rectale}, and \textit{Methanobrevibacter smithii}. These species represent the three main phyla in the human gut: Bacteroidetes, accounting for 17-60\% of the total biomass; Firmicutes, 35-80\% of the biomass; and Euryarchaeota the bulk of the remainder\cite{Shoaie2013}. The system’s main product, butyrate, is of specific interest due to its role in sustaining human health. Butyrate provides energy to colonocytes, affects overall energy homeostasis, and inhibits histone deacetylase, which is an enzyme that directly affects colorectal cancer\cite{Shoaie2013}. Along with butyrate, other notable intermediates and products in this system include acetate, propionate, glutamine, carbon dioxide (CO$_2$), hydrogen (H$_2$), and methane (CH$_4$). Acetate, propionate, and butyrate, which are short chain fatty acids, are absorbed in the gut’s epithelial cells and regulate an individual’s immune system and metabolism\cite{Shoaie2013}. Individually, acetate acts as a substrate for cholesterol synthesis and lipogenesis \cite{Shoaie2013}; propionate regulates gluconeogenesis and cholesterol synthesis\cite{Shoaie2013}; glutamine fuels the metabolism and maintains the intestinal barrier\cite{Kim2017}; and the gases CO$_2$, H$_2$, and CH$_4$ are products of bacterial fermentation that can cause intestinal discomfort in excess\cite{Scaldaferri2013}. As for the microorganisms themselves, \textit{M. smithii} removes hydrogen gas, which affects bacterial fermentation and energy gathering, and produces methane gas\cite{Shoaie2013}; \textit{E. rectale} produces butyrate, which is beneficial to the gut’s epithelial cells\cite{Shoaie2013}; and \textit{B. thetaiotaomicron} utilizes dietary polysaccharides and indirectly facilitates butyrate production with its outputs \cite{Ji2015}. 

\begin{figure}[ht]
    \centering
    \includegraphics[scale = 0.3]{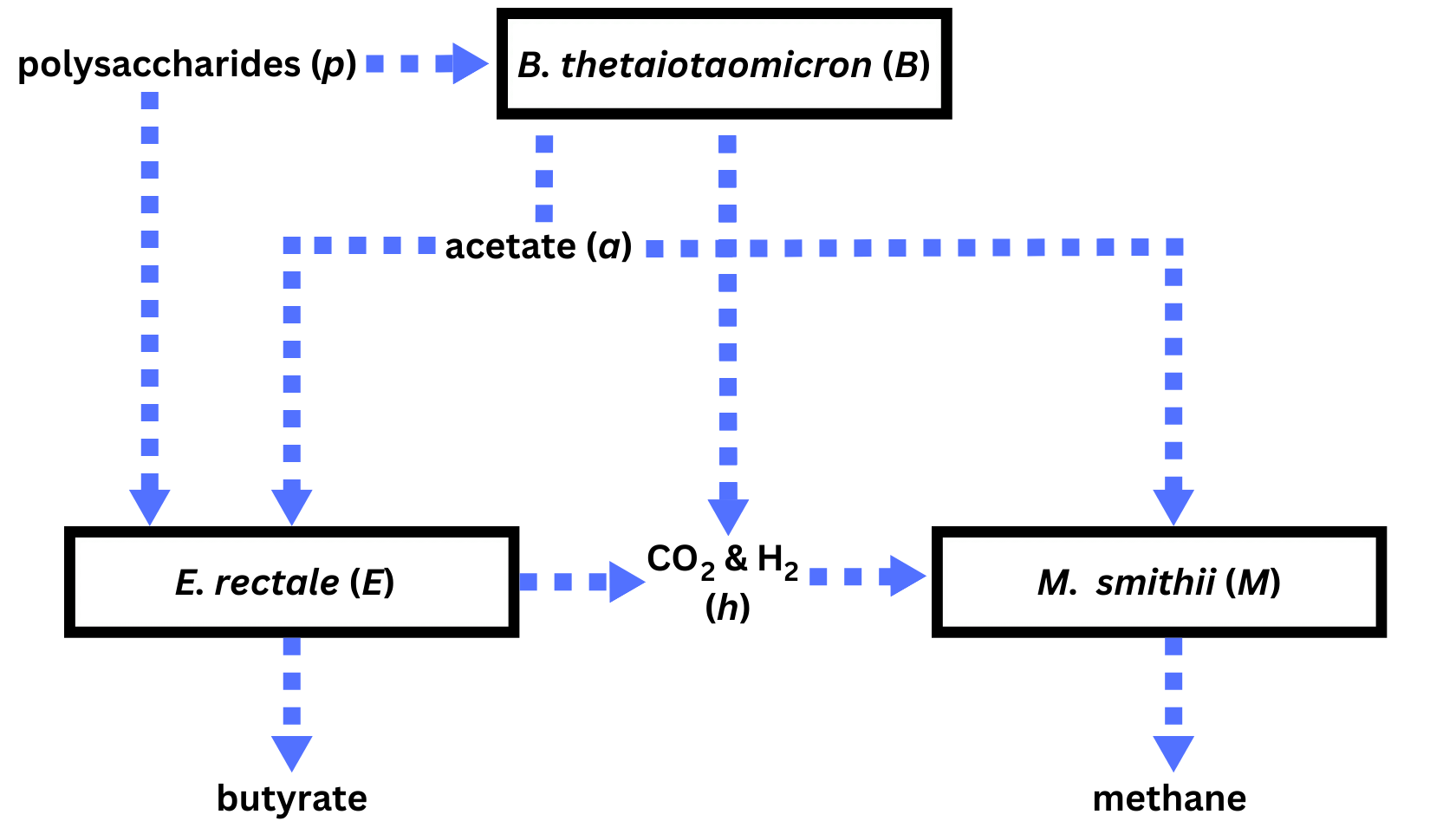}
    \caption[Graphical representation of the interactions of the three species, \textit{B. thetaiotaomicron}, \textit{E. rectale}, and \textit{M. smithii}, and their metabolites.]{Graphical representation of the interactions of the three species, \textit{B. thetaiotaomicron}, \textit{E. rectale}, and \textit{M. smithii}, and their metabolites. Each microorganism or substance given a letter is included in the mathematical model equations. This figure was developed from information presented in Ji \& Nielsen 2015 \cite{Ji2015}, Shoaie et al. 2013\cite{Shoaie2013}, and Adamberg et al. 2014 \cite{Adamberg2014}. We emphasize that \textit{B. thetaioaomicron}'s products acetate and the gases CO$_2$ and H$_2$ cannot be produced without the presence of polysaccharides.}
    \label{3species}
\end{figure}

In our model, we represent the interactions of these three short-chain fatty acid producing/utilizing gut microorganisms (see Figure \ref{3species}) and their metabolites. \textit{B. thetaiotaomicron} is an abundant bacterial species in the human gut microbiome whose main function is the utilization of polysaccharides \cite{Xu2003,Ji2015}. Through polysaccharide degradation, \textit{B. thetaiotaomicron} contributes to the overall ecosystem diversity in the colon, which is its regular environment \cite{Adamberg2014}. \textit{B. thetaiotaomicron} can survive solely on the uptake of carbon-rich polysaccharides\cite{Martens2011}; however, its growth is enhanced in the presence of inorganic ammonia due to inorganic ammonia's contribution of nitrogen to \textit{B. thetaiotaomicron}'s metabolism \cite{Glass1980}. Through the utilization of inorganic ammonia, \textit{B. thetaiotaomicron} can synthesize all amino acids that are essential to human health \cite{Ji2015}, which makes this bacteria a key interest in our study.

When both \textit{E. rectale} and \textit{B. thetaiotaomicron} are present in an environment, \textit{B. thetaiotaomicron} up-regulates gene expression for starch utilization and the degradation of specific glycans that \textit{E. rectale} is unable to utilize. Simultaneously, \textit{E. rectale} down-regulates the genes associated with glycan degradation even though it cannot grow efficiently without a carbohydrate source. Previous research on the interactions of these two species suggests that the presence of \textit{B. thetaiotaomicron} enhances \textit{E. rectale}'s ability to uptake nutrients\cite{ Mahowald2009}. \textit{E. rectale} shifts from uptaking polysaccharides to utilizing amino acids, such as glutamine, when \textit{B. thetaiotaomicron} is present\cite{Ji2015}. This reciprocal response suggests that these two gut microorganisms may adapt their metabolic strategies in the presence of one another to reduce competition for the same nutrients and optimize their use of available resources\cite{Martens2011}.

\textit{M. smithii}, which is one of the main methanogenic archaeon in the human gut, improves the productivity of carbohydrate metabolism by utilizing H$_2$ from \textit{E. rectale} and formate from \textit{B. thetaiotaomicron} to produce methane gas. This process prevents the environment from becoming too saturated with \textit{B. thetaiotaomicon} and \textit{E. rectale}'s by-products, which consequently improves carbohydrate metabolism. Additionally, \textit{M. smithii} removing H$_2$ in this environment allows for \textit{B. thetaiotaomicron} to generate NAD$^+$, which is used for glycolysis, a fundamental process in producing cellular energy \cite{ Mahowald2009}. 

\textit{M. smithii} is known to use substrates, including hydrogen and carbon dioxide, for methane production \cite{Martens2011}. The availability of amino acids produced by \textit{B. thetaiotaomicron} can serve as alternative carbon and energy sources, which may compete with these traditional substrates\cite{Martens2011}. If \textit{B. thetaiotaomicron}-produced amino acids are abundant, \textit{M. smithii} may favor their utilization over other substrates, potentially reducing methane production\cite{Martens2011}.

The information known about these three species' interactions is translated into a graphical representation in Figure \ref{3species}. This schematic highlights the competition among these species, specifically between \textit{E. rectale} and \textit{M. smithii} and between \textit{B. thetaiotaomicron} and \textit{E. rectale}.

The results of our modeling effort suggest the efficacy of a relatively simple ODE approach in understanding and potentially predicting the dynamics of important subsystems in the larger human gut ecosystem.

\section*{Results}
Using our mathematical model of \textit{B. thetaiotaomicron}, \textit{E. rectale}, and \textit{M. smithii} and their metabolites in equations (\ref{subequations1}) and (\ref{subequations2}), we present numerical results, estimation of key parameters, and sensitivity analysis. The code used to generate our results can be found at our GitHub page:  \url{https://github.com/Melissa3248/gut_microbiota}.

\subsection*{Numerical Results}

\begin{figure}[ht]
    \centering
    \includegraphics[scale = 0.10]{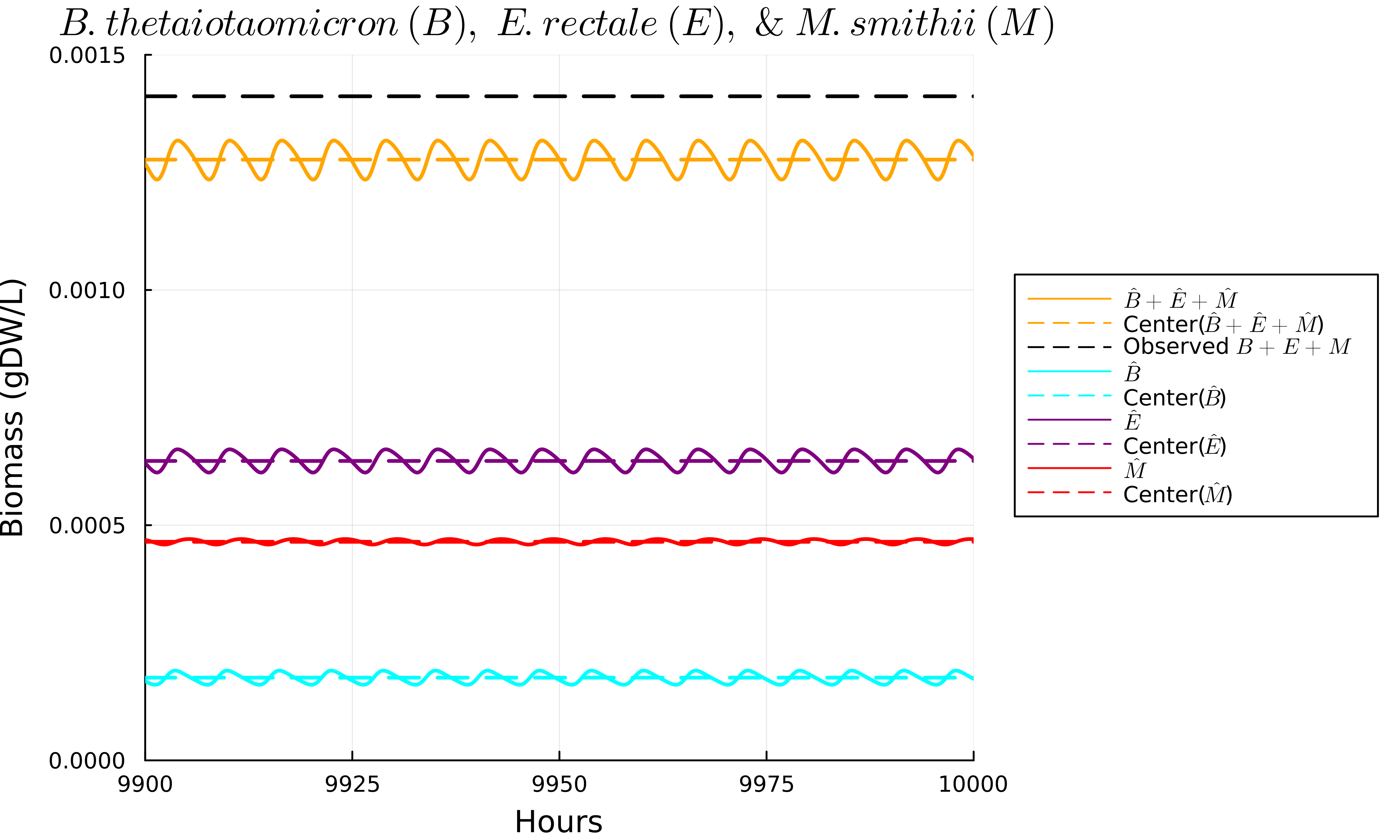}
    \caption{Plot of the solutions to equations (\ref{B}), (\ref{E}), and (\ref{M}) using the observed data in Supplementary Table 1 over the time interval 9,900 to 10,000 hours. Based on Supplementary Table 1, the sum of the three species' biomass should be 0.001412 gDW. The middle of the solutions to the ODEs using the parameter estimates in Table \ref{paramEstimFull} for all three species sums to 0.001277 gDW, resulting in an error of $-$0.000135 gDW, or a $-9.5$\% error.}
    \label{BEMplot}
\end{figure}

\begin{figure}[ht]
    \centering
    \includegraphics[scale = 0.10]{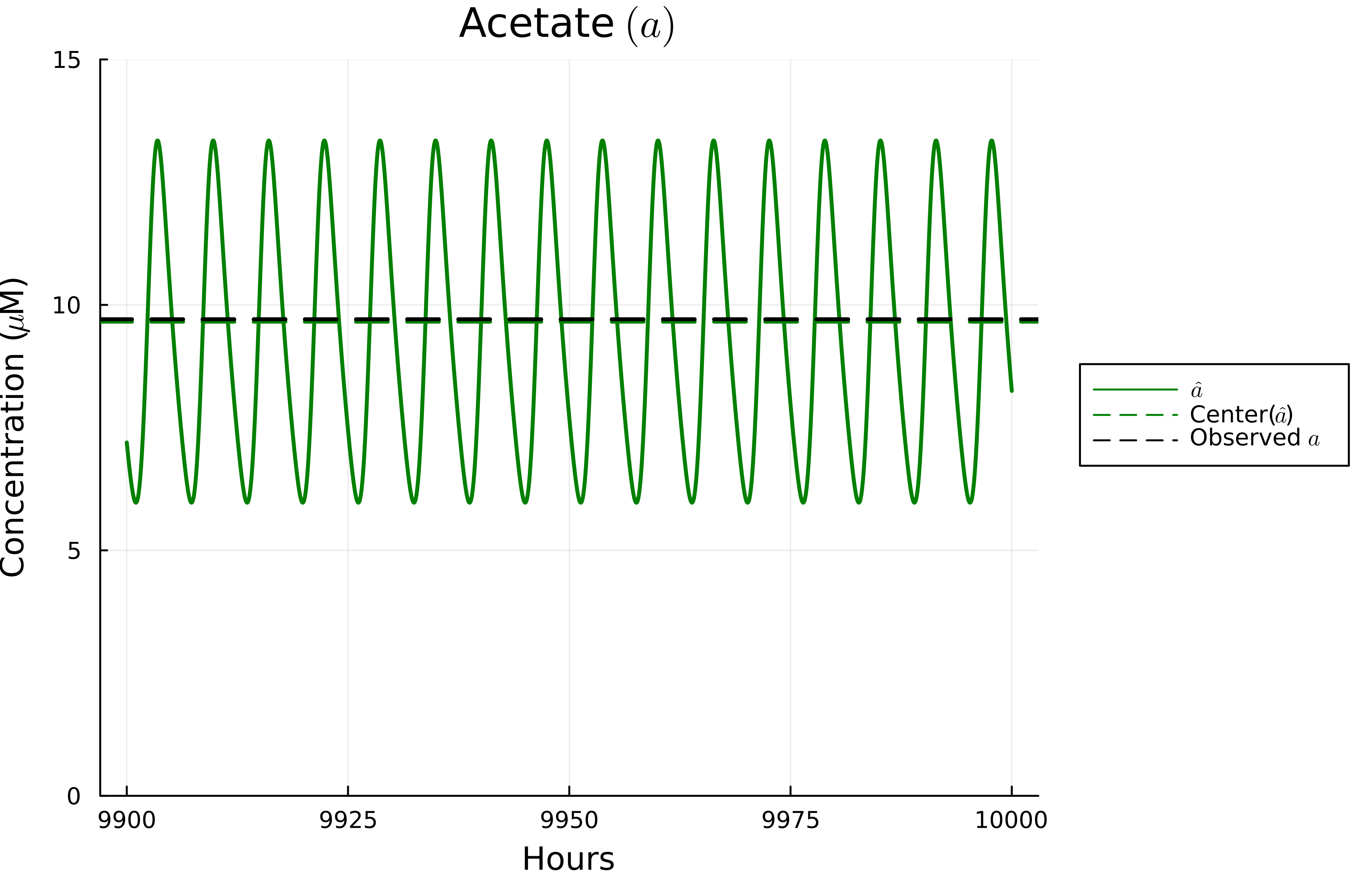}
    \caption{Plot of the solutions to equation (\ref{a}) using the observed data in Supplementary Table 1 over the time interval 9,900 to 10,000 hours. Based on Supplementary Table 1, the center of acetate's oscillations should be 9.71 $\mu$M. The middle concentration of the ODE solutions for acetate is 9.66 $\mu$M, resulting in a -0.05 $\mu$M error, or a $-$0.5\% error.}
    \label{Acetateplot}
\end{figure}

\begin{figure}[ht]
    \centering
    \includegraphics[scale = 0.10]{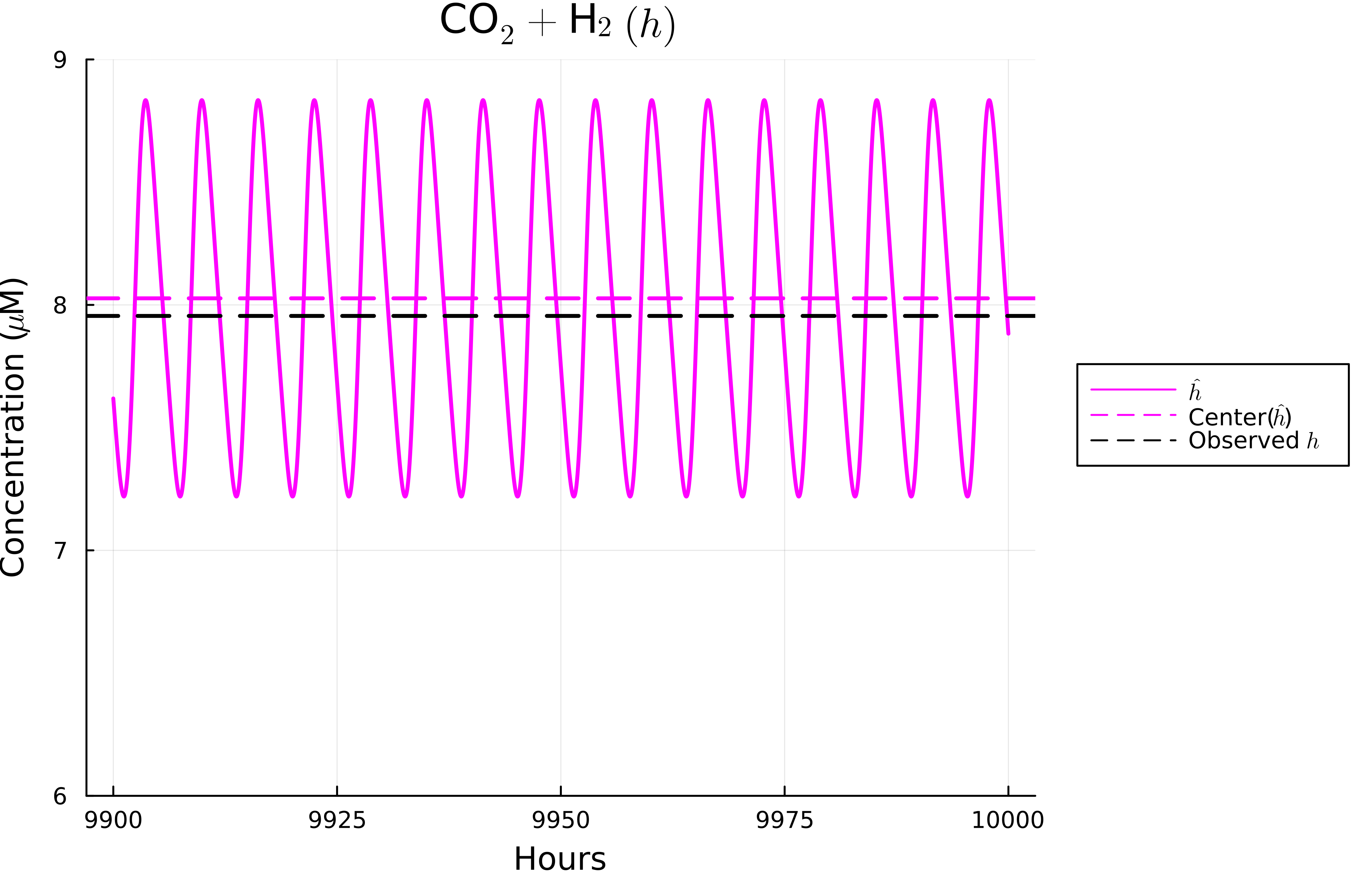}
    \caption{Plot of the solutions to equation (\ref{h}) using the observed data in Supplementary Table 1 over the time interval 9,900 to 10,000 hours. Based on Supplementary Table 1, the center of CO$_2$ and H$_2$'s oscillations should be roughly 7.96 $\mu$M. The middle concentration of the ODE solutions for CO$_2$ and H$_2$ is 8.03 $\mu$M, resulting in a $0.07$ $\mu$M error, or a 0.9\% error.}
    \label{CO2H2plot}
\end{figure}

\begin{figure}[ht]
    \centering
    \includegraphics[scale = 0.10]{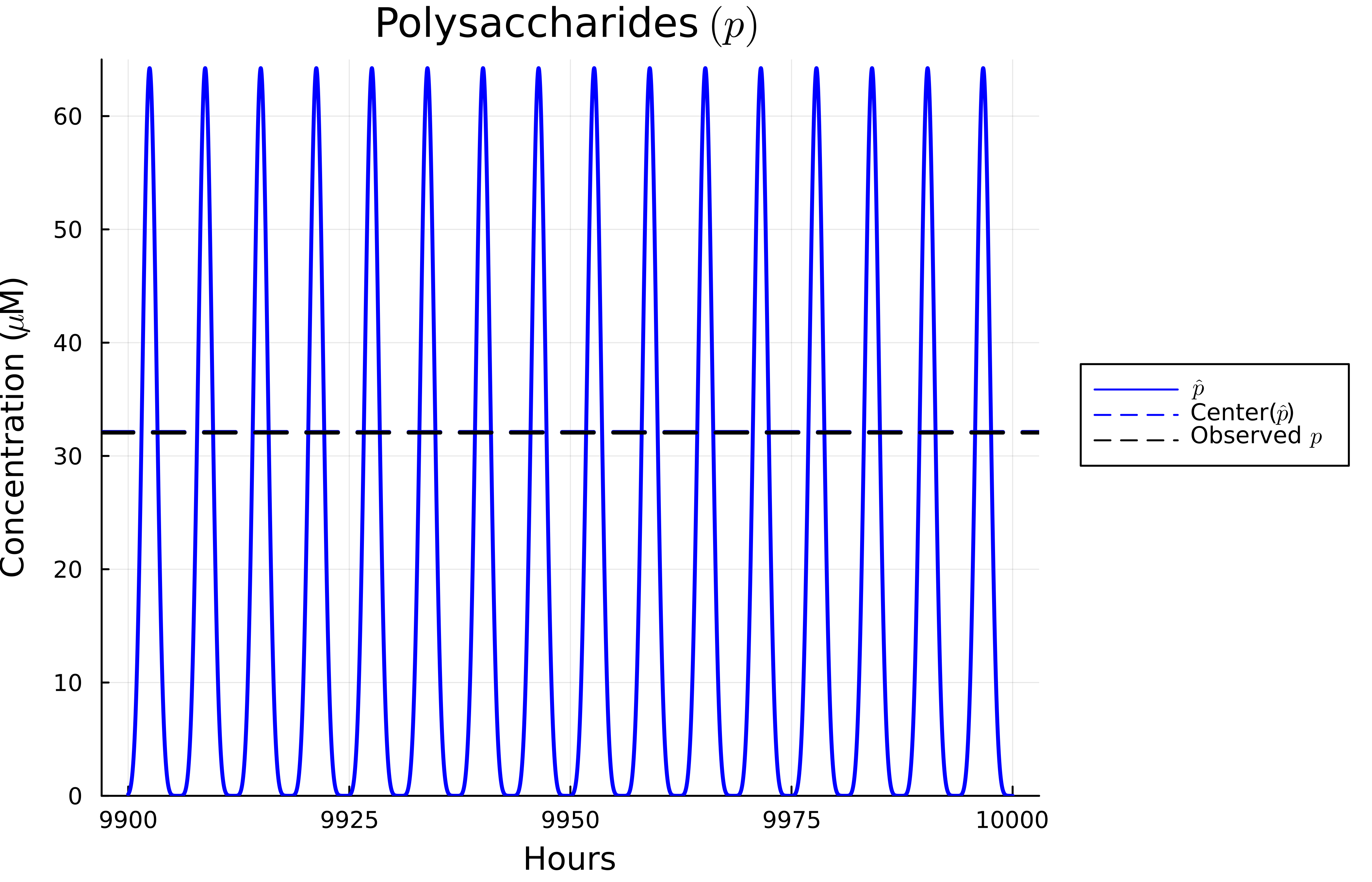}
    \caption{Plot of the solutions to equation (\ref{p}) using the observed data in Supplementary Table 1 over the time interval 9,900 to 10,000 hours. Based on Supplementary Table 1, the center of polysaccharides' oscillations should be roughly 32.06 $\mu$M. The middle concentration of the ODE solutions for polysaccharides is 32.11 $\mu$M, resulting in a 0.05 $\mu$M error, or a 0.2\% error.}
    \label{Polyplot}
\end{figure}

Using the set of parameter estimates in Table \ref{paramEstimFull}, the solutions to our model are given for the dynamics of \textit{B. thetaiotaomicron}, \textit{M. smithii}, and \textit{E. rectale} in Figure \ref{BEMplot}, for acetate in Figure \ref{Acetateplot}, for CO$_2$ and H$_2$ in Figure \ref{CO2H2plot}, and for polysaccharides in Figure \ref{Polyplot}. These plots show the solutions to our system of ODEs in equations (\ref{subequations1}) and (\ref{subequations2}) from 9,900 to 10,000 hours. This time range was selected in order to allow for a sufficient amount of time to pass in order for the system to converge to a regular oscillation. The center values of the oscillations are represented by dotted lines in each plot. We note in Figure \ref{BEMplot} that the mean of three bacterial species was fitted to match the experimental biomass sum instead of individually. This choice was made given the data availability; the data available reported the sum of the biomass. We note that there can be many different combinations of individual biomass for the three species that lead to this specific sum. We emphasize that the fact that we found a set of parameters that allow for the species to coexist long term provides some validation of our model since we know these species coexist naturally in the gut.

In Figures \ref{Acetateplot}, \ref{CO2H2plot}, and \ref{Polyplot}, the data values given in Supplementary Table 1 are superimposed on their respective plots as dashed lines in order to provide a visual representation of the error between the mean of the regular oscillation and the observed data. Despite some error in the center value of the ODE solutions compared to the data, the observed data values for acetate, polysaccharides, and the gases CO$_2$ and H$_2$ are contained in the oscillation range of the ODEs' numerical solutions, so we conclude that our parameter estimates sufficiently fit the data.  Thus, the relatively simple model we have presented here is consistent with the data in our literature review. Further details on our parameter estimation process can be found in Supplementary Information Appendix B.

\subsection*{Sensitivity Analysis}
Based on the fitted parameter values and the available data from the literature, we analyzed our model's sensitivity to its parameters.
In order to identify the parameters in the model that have the greatest effect on the model output, we conducted sensitivity analysis on our model by computing the first- and total-order effects based on Sobol' indices.

\begin{figure}[h!]
    \centering
    \includegraphics[scale = 0.45]{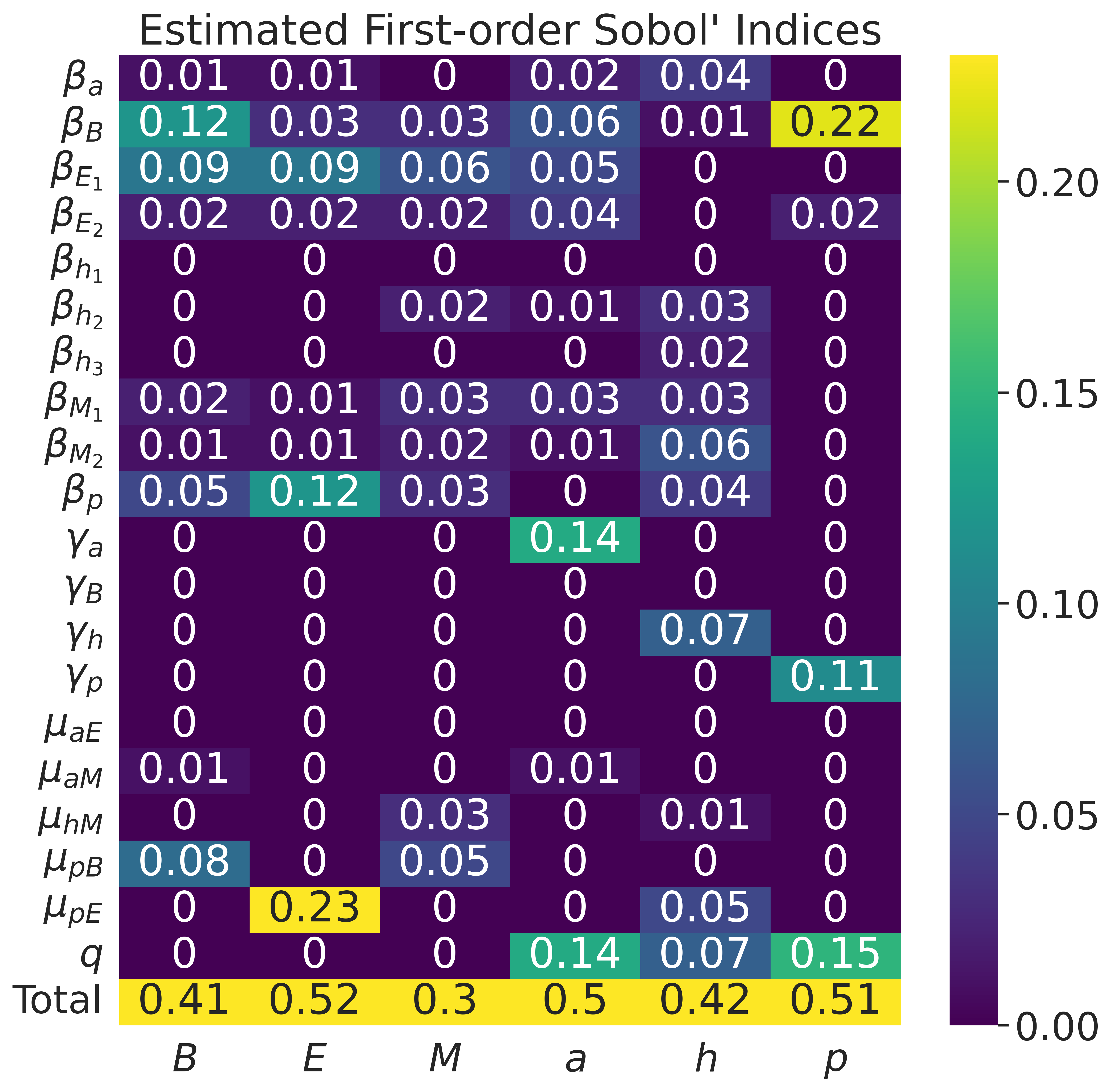}
    \caption{Heatmap of model parameters' estimated first-order Sobol' indices for each output, $B$, $E$, $M$, $a$, $h$, and $p$, using $N=2^{19}$ simulations.}
    \label{SiTable}
\end{figure}

\begin{figure}[h!]
    \centering
    \includegraphics[scale = 0.45]{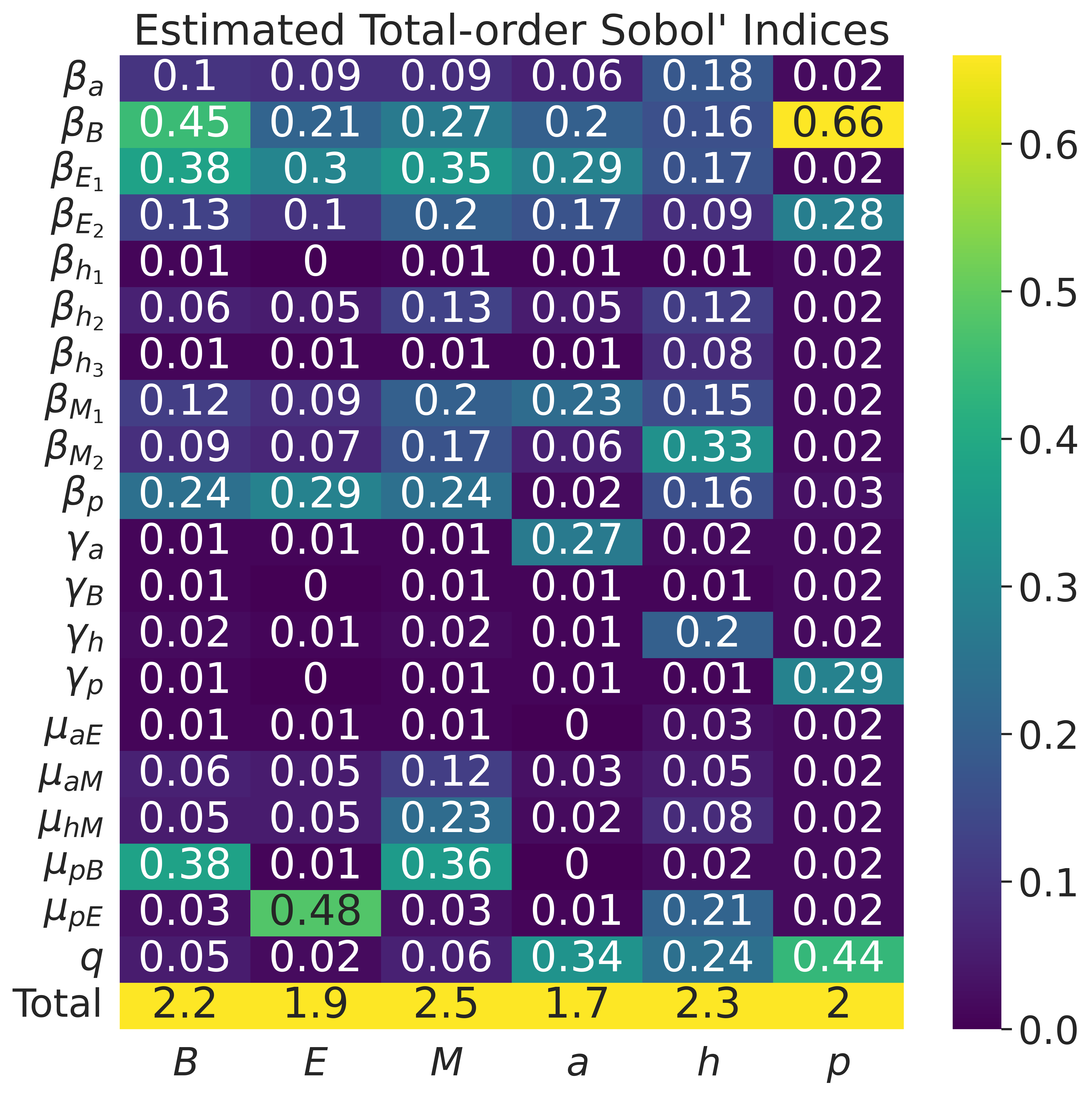}
    \caption{Heatmap of model parameters' estimated total-order Sobol' indices for each output, $B$, $E$, $M$, $a$, $h$, and $p$, using $N=2^{19}$ simulations.}
    \label{STiTable}
\end{figure}

Sensitivity analysis can be defined as the study of how uncertainty in the model input propagates into uncertainty in the model output\cite{Saltelli2008}. Once parameters are estimated, sensitivity analysis can be used to identify the driving parameters of the system that contribute to the most variability in the model's output. Sensitivity analysis tests the robustness of the model, identifies if the model relies on weak assumptions, and allows for model simplification\cite{Saltelli2008}. The results of our sensitivity analysis using the methods described in the section titled "Sensitivity Analysis Methods" are given in Figures \ref{SiTable} and \ref{STiTable}.

We obtained first- and total-order Sobol' indices results, shown in Figures \ref{SiTable} and \ref{STiTable}. Based on Figure \ref{SiTable} of the first-order Sobol' indices, $\beta_B$ is an influential parameter for \textit{B. thetaiotaomicron}'s output variance; $\beta_p$ and $\mu_{pE}$ for \textit{E. rectale}'s variance, $\gamma_a$ and $q$ for acetate's variance, and $q$ and $\beta_B$ for polysaccharide's variance. 

Based on Figure \ref{STiTable} of the total-order Sobol' indices, we can qualitatively see that parameters that we would biologically expect to be relevant to the variance of corresponding output are in fact relevant. Specifically, $\beta_B$ has a higher order effect on the output variance of polysaccharide's $(p)$, the consumption term $\mu_{pE}$ has a higher order effect on the output variance of \textit{E. rectale}, $\beta_B$ has a higher order effect on the output variance of \textit{B. thetaiotaomicron}, etc. 
Based on the changes in the estimated indices from first-order to total-order indices, there appears to be a higher-order interaction among model parameters for \textit{E. rectale}'s, \textit{M. smithii}'s, acetate's, CO$_2$ and H$_2$'s, and polysaccharides' output variance. We note that observing Sobol' indices whose sum is greater than 1, as seen in Figure \ref{STiTable}, is evidence of potentially correlated inputs in the model \cite{Iooss2019}. 


\section*{Discussion}

We suggest and illustrate that a mathematical representation using inflow and outflow terms, such as those used in models of chemostats, is a natural way to capture the effects of the inflow and outflow in the gut on its microbial dynamics. The use of inflow and outflow terms in our approach differentiates our model from many other dynamical systems representations, such as batch models.  One difference is that in a chemostat model with inflows and outflows, persistence of all species is possible, whereas this may would usually not be the case in batch models, even with otherwise identical representations of the cellular and chemical interactions.

The three species in our model, \textit{B. thetaiotaomicron}, \textit{E. rectale}, and \textit{M. smithii}, play an important role in polysaccharide degradation and the production of butyrate, which both aid in the human gut’s ability to absorb nutrients through the epithelial cells\cite{Shoaie2013}. The system of the three microbial species has been considered in previous works\cite{Shoaie2013,Ji2015}, which largely informed our knowledge of this system. By creating a mathematical model based on the interactions of these species, we have analyzed their interactions and identified aspects of this system that should be further explored through empirical investigation. 

When dealing with observed phenomena, it is often the case that several different mathematical representations can mimic those observations. Models, even simplified and abstracted models like ours, contain critical assumptions concerning the relationships between constituent parts. The implications of these assumptions are at least somewhat illuminated by parameter sensitivity analyses, which can identify the most impactful relationships. 

Due to the limited availability of data for a more rigorous parameter estimation, our sensitivity analysis took on additional importance. Through the results of the sensitivity analysis using first- and total-order Sobol’ indices, we more narrowly identified specific links in the microbial food web that would be fruitful targets for additional empirical work. Specifically, we identified the growth rate coefficients $\beta_B$, $\beta_{E_1}$, $\beta_{E_2}$, $\beta_{M_1}$, $\beta_{M_2}$, $\beta_p$, $\mu_{pB}$, $\mu_{pE}$ and the scaled volumetric flow $q$ as being largely significant in contributing to the variance of the model output, including higher-order interactions among these parameters.  With these results, we suggest that estimates of these significant parameters be obtained through laboratory experimentation in order to capture these values to a higher degree of precision and accuracy.

The significant growth rate coefficients ($\beta$ parameters) correspond to the growth rates of the three microorganisms when supported on a medium containing specific nutrients. Experiments should focus on cultivating these microorganisms in isolation in germ-free mice with only one nutrient \cite{Shoaie2013}. Despite the fact that these microorganisms are able to be cultured in isolation of other microorganisms, they may not be able to be sustained on one single nutrient, but rather require the presence of additional substrates. In this case, nonlinear effects from these secondary nutrients would factor in to the resulting estimates of the growth parameters. Table \ref{Experiment} provides a general outline of the the specific nutrient and microorganism necessary to estimate each parameter. For example, the growth rate coefficient for \textit{B. thetaiotaomicron} ($\beta_B$) can be experimentally estimated by cultivating \textit{B. thetaiotaomicron} in a medium containing only polysaccharides. Additionally, experiments approximating the rate of flow of the digestive system's fluids would largely inform the estimate for the true value of the volumetric flow rate parameter $q$.

\begin{table}[ht]
\centering
\begin{tabular}{ccc}\toprule
Parameter & Microorganism & Nutrient \\
\midrule
$\beta_{B}$ & \textit{B. thetaiotaomicron} & polysaccharides \\
$\beta_{E_1}$ & \textit{E. rectale} & acetate \\ 
$\beta_{E_2}$ & \textit{E. rectale} & polysaccharides \\ 
$\beta_{M_1}$ & \textit{M. smithii} & acetate \\
$\beta_{M_2}$ & \textit{M. smithii} & CO$_2$ and H$_2$ \\
$\mu_{pB}$ & \textit{B. thetaiotaomicron} & polysaccharides \\
$\mu_{pE}$ & \textit{E. rectale} & polysaccharides \\ \bottomrule
\end{tabular}
\caption[Table of microorganism growth rate parameters suggested to be estimated experimentally.]{Table of microorganism growth rate parameters suggested to be estimated experimentally.}
\label{Experiment}
\end{table}
 
 Due to the possibility of correlation among model parameters, variance-based sensitivity analyses specifically for correlated parameters should be explored and applied to this system, such as the methods discussed in Iooss and Prieur 2019\cite{Iooss2019} and Rabitz 2010\cite{Rabitz2010}. In regards to specifying the prior distributions on parameters, further analyses should include testing differing prior parameter distributions in order to compute the Sobol' indices, especially to explore the amount of variability in results based on the chosen parameter distributions\cite{Saltelli2008}.

Our main suggestion for future work is to collect more complete longitudinal data on this biological system, including for all three species \textit{B. thetaiotaomicron}, \textit{E. rectale}, and \textit{M. smithii} and its relevant substrates. With this experimental data, more precise estimates of the model parameters can be achieved. 
With this approach, particular attention can be paid to estimating the parameters that were identified as sensitive by the Sobol' indices. 

Additionally, further extensions of our model may include relaxing some of the volumetric flow assumptions. Specifically, the rate of volumetric flow through the gut can be generalized to reflect aspects of the natural flow of the gut, such as periodically restricted flow, and the framework of a single vessel representation can be extended to consider additional compartments to mimic the system of digestive organs in the human body. More details about a chemostat framework and existing theory can be found in Smith and Waltman 2008\cite{SmithHalL.andWaltman2008} and Harmand et al. 2017\cite{Harmand2017}. Future implementations of a similar model can account for absorption rates of additional substrates within the gut, 
 particularly amino acids. With these potential improvements of our baseline model, additional aspects about the dynamics of this biological system can be uncovered, and these improvements to our model could fuel further research directions related to this system.

Our model has relevance to human microbiome studies as it can be used to test the significance of clinical findings.  For example, microbiome studies for human autoimmune disease, such as multiple sclerosis, have shown modulation of butyrate and methane-producing gut bacteria \cite{Jangi2016}.   This can be incorporated into our modeling framework in combination with experimental studies to determine the mechanism through which gut bacterial modulate disease. 

\section*{Methods}

The key assumption underlying our mathematical model is that the human gut can be represented as a chemostat. A chemostat is a laboratory device used in the simulation and ecological study of populations, which provides an idealized representation of naturally occurring phenomena and has a rich history of mathematical representation. Though the conditions of a chemostat are simplified and controlled in a laboratory setting, a chemostat can be useful in the study of population dynamics and the underlying mechanisms of interactions among populations. Using aspects of a simple chemostat model is a first step in developing an initial theoretical framework, which can then be refined and extended. 

In developing an ODE model to represent our chosen microbiota subsystem, we assume that the human gut has inflows and outflows in a manner similar to chemostat-like models.
We utilize ODE-based dynamical systems modeling to track the changes in biomass of the three prevalent microorganisms, \textit{B. thetaiotaomicron}, \textit{E. rectale}, and \textit{M. smithii}, as well as their chemical inputs, intermediates, and byproducts, with the goal of providing a better understanding of their interactions within this subsystem.

In order to supplement the knowledge gained from transcriptomic analysis and GEMs, we incorporate this information into our own mathematical interpretation of this small-scale system using ODEs. Throughout our model explanation, we present information learned through previous investigations of \textit{B. thetaiotaomicron}, \textit{E. rectale}, and \textit{M. smithii} that we utilized in creating our mathematical model.

In constructing a first model, we assume that the contents of the vessel are well-mixed, the rate at which liquid enters the system equals the rate at which the well-stirred contents leave the compartment, and that some significant factors potentially affecting growth, such as temperature, are held constant\cite{SmithHalL.andWaltman2008}. 

For specificity we assume that microorganisms grow at a rate following the Monod form 
\begin{equation}
\beta_X\left(\frac{u}{u+\gamma}\right)X  \label{monod_form}  
\end{equation}
\noindent where $\beta_X$ is the maximum birth rate of population $X$, $u$ is the concentration of the nutrient population on which $X$'s growth depends, $\gamma$ is the Michaelis-Menten constant, and $X$ is the concentration of the microorganism\cite{Monod1949}. This form is commonly used, especially for a first effort.  Other functional forms that could be used include Hill functions\cite{Stefan2013}, which are generalizations of the Monod form, and S-forms\cite{Voit2000}.  The accompanying large increase in the number of parameters, and lack of data to capture the detail these parameters provide, means that they are less appropriate for our effort here than the Monod form.

\subsection*{Model Variables}

The variables in our model for the microbial and chemical species depend only on one independent variable, time, denoted by $t$.  These dependent variables are

\begin{itemize}
    \item $B$ for {\em B. thetaiotaomicron},
    \item $E$ for {\em E. rectale},
    \item $M$ for {\em M. smithii},
    \item $a$ for acetate,
    \item $h$ for CO$_2$ and H$_2$,
    \item $p$ for polysaccharides.
\end{itemize}

\subsection*{Model Equations}

The equations for the three microbial species are 
\begin{subequations} \label{subequations1}
\begin{align}
\frac{dB}{dt} & = \beta_B\Psi_{\gamma_p}(p)B - qB \label{B}, \\ 
\frac{dE}{dt} & = \left[\beta_{E_1}\Psi_{\gamma_a}(a) +  \beta_{E_2}\left(1-\Psi_{\gamma_B}(B)\right)\Psi_{\gamma_p}(p)\right]E - qE \label{E} ,\\ 
\frac{dM}{dt} & = \left[\beta_{M_1}\Psi_{\gamma_a}(a) + \beta_{M_2}\Psi_{\gamma_h}(h)\right]M - qM \label{M}.
 \end{align}
\end{subequations}

The equations for the three chemical species are

\begin{subequations} \label{subequations2}
\begin{align}
\frac{da}{dt} &= \beta_a\Psi_{\gamma_p}(p)B-qa-\mu_{aE}\Psi_{\gamma_a}(a)E - \mu_{aM}\Psi_{\gamma_a}(a)M \label{a}, \\
\frac{dh}{dt} &= \beta_{h_1}\Psi_{\gamma_a}(a)E + \beta_{h_2}\Psi_{\gamma_p}(p)B + \beta_{h_3}(1-\Psi_{\gamma_B}(B)  )\Psi_{\gamma_p}(p)E -qh-\mu_{hM}\Psi_{\gamma_h}(h)M \label{h}, \\
\frac{dp}{dt} &= \beta_pq(\textrm{cos}(t)+1)^3 - qp - \mu_{pB}\Psi_{\gamma_p}(p)B - \mu_{pE}(1 -\Psi_{\gamma_B}(B))\Psi_{\gamma_p}(p)E \label{p}.
\end{align}
\end{subequations}

The $\Psi$ and $q$ terms in these equations are defined by
\begin{align}
    \Psi_{\gamma}(u)&=\frac{u}{u+\gamma}, \label{monod} \\
    q&=\frac{V}{Q} \label{q}.
\end{align}

In our model we assume that there is a high enough rate of turnover of fluids in the human gut such that these microorganisms are almost always flushed out of the system before their life expectancy, so death terms are neglected in these three equations. All three microorganism equations thus have the general format: $$\Delta_i = P_{ij}- F_i,$$ where $\Delta_i$ is the rate of change of biomass for microorganism $i$, $P_{ij}$ is the rate at which microorganism $i$ proliferates based on the availability of substance $j$, and $F_i$ is the rate at which microorganism $i$ is flushed out of the gut. In all three microorganism equations, the term $F_i$ is the biomass of the given population multiplied by the constant $q$. The fixed quantity $q$ is interpreted as the rate at which the contents of the gut leave the system as expressed in equation (\ref{q}), where $V$ is the volume of the gut and $Q$ is the rate of volumetric flow within the gut. 

In order for \textit{E. rectale} to grow in biomass, acetate or polysaccharides need to be present in the ecosystem\cite{Shoaie2013}. We assume that \textit{E. rectale} has different maximum growth rates depending on each nutrient, leading us to split $\beta_E$ into three different, related constants. Because \textit{E. rectale} shifts to uptaking inorganic ammonia when \textit{B. thetaiotaomicron} is present, we included the term $\big(1-\Psi_{\gamma_B}(B)\big)$ to reflect this shift. 
\textit{M. smithii} depends on the presence of acetate and the gases CO$_2$ and H$_2$, which are both incorporated in the standard Monod form from equation (\ref{monod_form}). Again, we assume that \textit{M. smithii} grows at different maximum growth rates in the presence of only one of these metabolites, which lead to the separation of $\beta_M$ into the constants $\beta_{M_1}$ and $\beta_{M_2}$.

The metabolite equations detail the rates of change over time for the intermediate substances produced and consumed by these three species, where the concentrations are tracked for acetate in equation (\ref{a}), CO$_2$ and H$_2$ in equation (\ref{h}), and polysaccharides in equation (\ref{p}). These concentrations depend on the rate at which these metabolites are produced by the microorganisms or enter into the system, the rate at which they are flushed out of the system, and the rate at which they are consumed by surrounding microorganisms. All three metabolite equations are constructed in the general format: 
$$\Delta_j = P_{ij} - F_j - C_{ij},$$
where $\Delta_j$ is the rate of change of metabolite $j$'s concentration, $P_{ij}$ is the rate at which metabolite $j$ is produced by microorganism $i$ or enters the system, $F_j$ is the rate at which metabolite $j$ is flushed out of the gut, and $C_{ij}$ is the rate at which the metabolite $j$ is consumed by microorganism $i$. In this $C_{ij}$ term for metabolite $j$ consumed by microorganism $i$, we have a yield constant $\mu_{ji}$, and in equations (\ref{a}) and (\ref{h}),  $\Psi_{\gamma_j}(j)X_i$ describes the contribution of metabolite $j$ to microorganism $i$'s biomass, where this biomass is denoted by $X_i$. For polysaccharides in equation (\ref{p}), $\left(1-\Psi_{\gamma_B}(B)\right)\Psi_{\gamma_j}(j)X_i$ describes the contribution of metabolite $j$ to microorganism $i$'s biomass depending on the biomass of \textit{B. thetaiotaomicron}.
 We include the term $\left(1-\Psi_{\gamma_B}(B)\right)$  in equations (\ref{h}) and (\ref{p}) to denote \textit{E. rectale}'s shift in utilization of polysaccharides and corresponding production of CO$_2$ and H$_2$ when \textit{B. thetaiotaomicron} is present.

Polysaccharides enter the human gut through diet, so we accounted for their addition to the gut through a sinusoidal function, $\big(\textrm{cos}(t)+1\big)^3$. This function attempts to account for the duration of time in between meals through the period of the curve. In addition, this function is defined to be a smooth curve to illustrate the gradual breakdown of food and release of nutrients in the gut. We note that the terms $\left(\text{cos}(t)+1\right)^3$ is one possible choice of many for representing how polysaccharides enter the system. Future work can explore different reasonable choices of this term and the effects on the system parameters and sensitivities.

The forms for the rate functions in equations (\ref{subequations1}) and (\ref{subequations2}) were chosen so that the system is mathematically conservative and thus meet the conservation criterion for a chemostat \cite{SmithHalL.andWaltman2008, Harmand2017}. Further reading of chemostat conservation principles and the construction of chemostat equations for competing species can be found in Harmand et al. 2017\cite{Harmand2017}. 

A mathematical model is a simplification and idealization. In our case, the three microbial species and their attendant chemical species are not an isolated system but rather part of a much larger ecosystem. In this larger ecosystem, other unaccounted for microorganisms and substrates interact with and affect the substrates and microorganisms we have considered. We note that it is often easier to investigate the dynamics and interactions of interest by focusing on and isolating the mechanisms within a relevant subsystem.  


\subsection*{Parameter Estimation Techniques}

In our model, many of the parameter values are unknown or cannot be determined experimentally. In order to estimate these parameters mathematically, we searched for data tracking the biomass changes of \textit{B. thetaiotaomicron}, \textit{E. rectale}, and \textit{M smithii} in order estimating these parameters. This system as its written in equations (\ref{subequations1}) and (\ref{subequations2}), however, does not account for all possible parameters that could potentially affect the fluctuations in biomass or substrate concentration. We provide a description of our data in Supplementary Information Appendix A. We emphasize that this data is limited in that it only contains information about one time point, which makes our parameter estimates highly uncertain.



By fitting the model parameters to the data shown in Supplementary Table 1, we obtained estimates of our model's parameters, which are shown in Table \ref{paramEstimFull}. We obtained this set of parameters by optimizing equation (\ref{eq:loss}) using the Nelder-Mead method. Optimization details can be found in Supplementary Information Appendix B.

\begin{equation}
    \text{MAPE}(\hat x_{t_1:t_2},x^\text{obs}) = \frac{1}{4}\sum_{i=1}^4  \left|\frac{\hat x_{t_1:t_2}-x^\text{obs}}{x^\text{obs}}\right|,   
    \label{eq:loss}
\end{equation}

\noindent where $\hat x_{t_1:t_2} = \left[\frac{1}{3}\left(\hat B_{t_1:t_2} + \hat E_{t_1:t_2} + \hat M_{t_1:t_2}\right), \hat a_{t_1:t_2}, \hat h_{t_1:t_2}, \hat  p_{t_1:t_2}\right]$,
$x^\text{obs} = \left[\frac{1}{3} (B^\text{obs} + E^\text{obs} + M^\text{obs}), a^\text{obs}, h^\text{obs}, p^\text{obs}\right]$, and $\hat x_{t_1:t_2}$ can be defined as $\hat x_{t_1:t_2}= \frac{1}{2}\big(\text{max}(x_{t_1:t_2}) + \text{min}(x_{t_1:t_2})\big)$. The matrix $x_{t_1:t_2} \in \mathbb{R}^{4\times(t_2-t_1)/\Delta t}$ \textcolor{blue} corresponds the the ODE solutions of equations \ref{subequations1} and \ref{subequations2} between times $t_1$ and $t_2$ with a $\Delta t$ time step. This choice of $\hat x_{t_1:t_2}$ allows us to measure the discrepancy between the observed data and the center of the oscillations of the ODE solutions between times $t_1$ and $t_2$. In our setting, we choose $t_1= 1400$ and $t_2 = 1500$ to allow the ODE solutions to approach a steady state.



\begin{table}[ht] 
\renewcommand*{\arraystretch}{1.4}
\centering
\begin{tabular}{ |c|c||c|c||c|c||c|c|  }
\hline
$\beta_a$ & 321,567$\frac{\mu\text{M}}{\text{gDW/L}*\text{hour}}$ & $\beta_{h_2}$ &  32,572$\frac{\mu\text{M}}{\text{gDW/L}*\text{hour}}$& $\gamma_B$  & 11 gDW/L & $\mu_{hM}$ & 3,310$\frac{\mu\text{M}}{\text{gDW/L}*\text{hour}}$ \\ \hline
$\beta_B$ & 1.277/hour & $\beta_{M_1}$ & 0.750/hour & $\gamma_h$ & 132 $\mu$M & $\mu_{pB}$ & 341,299$\frac{\mu\text{M}}{\text{gDW/L}*\text{hour}}$ \\ \hline
$\beta_{E_1}$ & 0.849/hour & $\beta_{M_2}$ & 0.456/hour & $\gamma_p$ & 412 $\mu$M& $\mu_{pE}$ & 5,243,172$\frac{\mu\text{M}}{\text{gDW/L}*\text{hour}}$ \\ \hline
$\beta_{E_2}$ & 0.523/hour & $\beta_p$ & 1,064 $\mu$M/hour & $\mu_{aE}$ & 49,647$\frac{\mu\text{M}}{\text{gDW/L}*\text{hour}}$ & $q$ & 0.054/hour\\ \hline
$\beta_{h_1}$ & 412$\frac{\mu\text{M}}{\text{gDW/L}*\text{hour}}$& $\gamma_a$ & 238 $\mu$M & $\mu_{aM}$ & 39,041$\frac{\mu\text{M}}{\text{gDW/L}*\text{hour}}$ & $\beta_{h_3}$ & 10,029$\frac{\mu\text{M}}{\text{gDW/L}*\text{hour}}$\\
\hline
\end{tabular}
\caption{Table of fitted parameter values for equations (\ref{subequations1}) and (\ref{subequations2}) based on the experimental data in Supplementary Table 1. In this table, we report rounded versions of the fitted parameter values.}
\label{paramEstimFull}
\end{table}

\subsection*{Sensitivity Analysis Methods} \label{sensitivityMethods}
 
We utilized global sensitivity analysis in order to identify the driving parameters of our model. In this framework, which utilizes Monte-Carlo (MC) methods, each parameter is assigned a probability density function (pdf) based on \textit{a priori} information known about the parameter. Samples are drawn from these probability density functions to evaluate the overall model output.

Because we are interested in detecting significant nonlinear interactions among the model parameters, implementations of global sensitivity analysis methods for linear relationships like the partial rank correlation coefficient (PRCC), the Pearson correlation coefficient (CC), and standardized regression coefficients (SRC) would not be useful in our case. Consequently, we employed the Sobol' method, a variance-based decomposition method\cite{Marino2009}. 

Sobol' indices are a MC variance-based approach to calculating all first-order and total-effects indices in a model with $k$ parameters. These sensitivity indices are computed based on model evaluations for $N$ simulations. In order to improve the sensitivity estimates, the values tested for each parameter in the model are drawn from quasi-random number generators. Though the use of quasi-random numbers from a given distribution is not necessary, we incorporate this approach into our computations. 

To compute the first- and total-order Sobol' indices for our reduced model parameters, we specified prior distributions for each parameter from which to draw samples. Parameter samples are drawn using lattice rule samples on the interval [$a_i$,$b_i$] for $i \in 1,\dots,20$ specified in Supplementary Table 3, using the QuasiMonteCarlo.jl \cite{QuasiMC} package. Additional details of our sensitivity analysis can be found in Supplementary Information Appendix B. 



\section*{Data availability}

The data used for our parameter estimation efforts is included in the Supplementary Information of this manuscript, and the Julia code used for our simulations is included in our GitHub page: \url{https://github.com/Melissa3248/gut_microbiota}.

\break 

\bibliography{ref}



\section*{Author contributions statement}

M.A. translated the biological system to a schematic, wrote the Julia code, and conducted the numerical simulations and sensitivity analyses.   M.A and B.A. created the mathematical models.  M.A. and A.M. related the models and underlying biology to each other.  B.A. conceived the project.  All authors contributed to the writing of the manuscript and reviewed the manuscript. 

\section*{Additional information}


\textbf{Competing interests} 

The authors declare no competing interests.

\appendix
\setcounter{table}{0}
\include{appendix}

\end{document}

%% file: appendix.tex











%
%
\thispagestyle{empty}

\appendix

\section*{Appendix A: Data} \label{Data}
Longitudinal data on all three species \textit{B. thetaiotaomicron}, \textit{E. rectale}, and \textit{M. smithii} is, to our knowledge, not openly available in the published literature. We were able to extract a single set of data points for our system including all three species and their relevant substrates from Shoaie et al\cite{Shoaie2013}.

Given the lack of available longitudinal data for this biological system, we assume that this data, shown in Supplementary Table \ref{data}, are the center values of the oscillations for each substrate or biomass quantity. One complication that arises from using this data is that the total biomass for all three species was experimentally measured as a single quantity, which is another factor that further contributes to our uncertainty in our full model parameter estimates. We fit the model parameters to produce numerical solutions with average biomass concentrations that roughly sum to the biomass quantity given in Supplementary Table \ref{data}.

\begin{table}[ht] 
\renewcommand*{\arraystretch}{1.4}
\centering
\begin{tabular}{ |c|c|  }
\hline
Polysaccharides ($\mu$M) & 32.06  \\ \hline
H$_2$ ($\mu$M)& 0\\ \hline
CO$_2$ ($\mu$M) & 7.96 \\ \hline
Acetate ($\mu$M)& 9.71 \\ \hline
Biomass (gDW/L) & 0.001412 \\ 
\hline
\end{tabular}
\caption[Table of experimental data presented in Shoaie et al. 2013\cite{Shoaie2013} of the microorganisms \textit{B. thetaiotaomicron}, \textit{E. rectale}, and \textit{M. smithii} and their substrates CO$_2$, H$_2$, acetate, and polysaccharides.]{Table of experimental data presented in Shoaie et al. 2013\cite{Shoaie2013} of the microorganisms \textit{B. thetaiotaomicron}, \textit{E. rectale}, and \textit{M. smithii} and their substrates CO$_2$, H$_2$, acetate, and polysaccharides. The biomass measurement is a combination of the biomass of the three microorganisms.}
\label{data}
\end{table}

\section*{Appendix B: Optimization and Sensitivity Analysis Details}\label{app:optim_sens}

\subsection*{B.1 Optimization}

Our optimization problem involves finding a parameter set $\theta\in \mathbb{R}^{20}$ such that we closely match the steady state solutions of our ODEs in equations (2) and (3) to the data in Supplementary Table \ref{data}, where $$\theta = [\beta_a, \beta_B, \beta_{E_1}, \beta_{E_2}, \beta_{h_1}, \beta_{h_2}, \beta_{h_3}, \beta_{M_1}, \beta_{M_2}, \beta_p, \gamma_a, \gamma_B, \gamma_h, \gamma_p, \mu_{aE}, \mu_{aM}, \mu_{hM}, \mu_{pB}, \mu_{pE}, q].$$

These parameters, though representative of physical quantities, do not have precise estimates in the literature to our knowledge. With a naive 
or randomly generated initial guess of this parameter set, this optimization will struggle to find a parameter set that produces a steady state solution that closely matches our observations. For this reason, we performed a pre-search of the parameter space to obtain an initial parameter set that produces roughly desireable results. This pre-search was informed by some general principles that we hypothesize are reasonable for our system: chemical reactions happen at a faster timescale compared to microbial replication, and our chemostat proxy of the gut has a slow fluid turnover rate. We list our initialization of $\theta$ in Supplementary Table \ref{paramInit}.

\begin{table}[ht] 
\renewcommand*{\arraystretch}{1.4}
\centering
\begin{tabular}{ |c|c||c|c||c|c||c|c|  }
\hline
$\beta_a$ & 3.0$\times 10^5$ & $\beta_{h_2}$ & 33,000 & $\gamma_B$ & 10 & $\mu_{hM}$ & 3,000 \\ \hline
$\beta_B$ & 1.2 & $\beta_{M_1}$ & 0.827 & $\gamma_h$ & 150 & $\mu_{pB}$ & 200,000 \\ \hline
$\beta_{E_1}$ & 0.8 & $\beta_{M_2}$ & 0.5 & $\gamma_p$ & 400 & $\mu_{pE}$ & 5.0$\times 10^6$ \\ \hline
$\beta_{E_2}$ & 0.6 & $\beta_p$ & 1,000 & $\mu_{aE}$ & 50,000 & $q$ & 0.05 \\ \hline
$\beta_{h_1}$ & 400 & $\gamma_a$ & 200 & $\mu_{aM}$ & 40,000 & $\beta_{h_3}$ & 10,000  \\
\hline
\end{tabular}
\caption[Table of initial values of our parameter set $\theta$ for equations (2) and (3).]{Table of initial values of our parameter set $\theta$ for equations (2) and (3).} \label{paramInit}
\end{table}

With this parameter initialization, we performed a constrained optimization of the MAPE score defined in (6) using the Nelder-Mead optimization method. We defined our lower-bound for our constraints on the parameter set to be $[0]^{20}$, and our upper-bound to be $[1\text{e}6, 10,10,10,1000,1\text{e}5,  1\text{e}5,10,10,1\text{e}5,1\text{e}4,1\text{e}4,1\text{e}4,1\text{e}4,1\text{e}6,1\text{e}6,1\text{e}5,1\text{e}8,1\text{e}8,10]$. The ordering of the parameter values in this vector is consistent with the order shown in $\theta$. Based on this optimization problem, we obtained a final set of parameters shown in Table 2.


\subsection*{B.2 Sensitivity Analysis}

We specified our prior distribution of the each of the parameters to follow a $\text{Uniform}(a_i,b_i)$ distribution, where we define each $a_i$ and $b_i$ for each parameter in Supplementary Table \ref{Sens_prior}.

\begin{table}
\renewcommand*{\arraystretch}{1.4}
\centering
\begin{tabular}{ |c|c|c|  }
\hline
Parameter & $a$ & $b$\\
\hline 
 $\beta_a$ & 160,783 & 1,607,834\\
  $\beta_B$ & 0.64 & 6.39\\
  $\beta_{E_1}$&   0.42 & 4.24\\
$\beta_{E_2}$ &  0.26 & 2.62\\
 $\beta_{h_1}$ & 201 & 2,060\\
  $\beta_{h_2}$ &16,286 & 162,858\\
$\beta_{h_3}$ &  5,015 & 50,145\\
$\beta_{M_1}$  &    0.37 & 3.75\\
$\beta_{M_2}$ &   0.23 & 2.28\\
 $\beta_p$ &   532 & 5,324\\
$\gamma_a$ &    119 & 1,192\\
$ \gamma_B$ &    5.62& 56\\
$\gamma_h$  &  66 &663 \\
$\gamma_p$ &   206 & 2,060\\
$\mu_{aE}$&  24,821 &248,240\\
$\mu_{aM}$ & 19,520 & 195,203\\
 $\mu_{hM}$ & 1,655 & 16,548\\
 $\mu_{pB}$ & 170,649& 1,706,495 \\
 $\mu_{pE}$  &   2.62e6 & 2.62e7\\
  $q$ &    0.027 & 0.27\\
\hline
\end{tabular}
\caption{Table of the lower and upper bounds defining the Uniform$(a_i,b_i)$ prior distributions for $i = 1,\dots,20$ indexing the parameter set $\theta$. These distributions were used to produce our sensitivity analysis results in Figures 6 and 7.}
\label{Sens_prior}
\end{table}

The upper and lower bounds were defined to be half and five times, respectively, the parameter estimate from Table 2. This choice of upper and lower bounds for our sensitivity analysis allowed us to remain in a stable region of the parameter space in order to obtain numerically stable ODE solutions for any given parameter sample. 

We explored a different specification of each upper and lower bound, where the upper bound was 10 times our parameter estimates from Table 2, and the lower bound was set to 0 to reflect our large uncertainty in our parameter estimation results. This specification of prior distribution resulted in a large number of unstable ODE solutions, and upon inspection, many of these had parameters or sets of related parameters that were extreme or mismatched ({\em e.g.,} a decay term with ten times greater rate, essentially replacing $\exp(-x)$ with $\exp(-10x)$ for some $x$, or a large consumption of polysaccharides with low growth of {\em B. thetaiotaomicron}). This issue was resolved when we narrowed our upper and lower bounds closer to our parameter estimate, but still with significant variation.

We produced first- and total-order Sobol' indices based on $2^{19}$ ODE solutions from time 0 to 250 hours. In order to obtain a vector output of the ODE solutions to compute the Sobol' indices, we average the ODE solutions beginning from hour 200 until 250, which resulted in a vector in $\mathbb{R}^{6}$ corresponding to our tracked quantities in our ODE, $B$, $E$, $M$, $a$, $h$, and $p$. We present our first-order Sobol' indices in Figure 6 and our total-order Sobol' indices in Figure 7.

